%% file: Template.tex
\documentclass{article}
\usepackage{spconf,amsmath,graphicx}
\usepackage{algorithmic}
\usepackage{graphicx, caption, subcaption}
\usepackage{textcomp}
\usepackage{xcolor}
\usepackage{bm}
\usepackage{multirow}
\usepackage{pgfplots}
\pgfplotsset{compat=1.5}
\usepackage{tikz}
\usepackage{balance}
\usepackage{url} 
\usepackage{tabularx}
\usepackage{float}
\usepackage{hyperref}

\AtBeginShipout{\AtBeginShipoutUpperLeft{%
		\put(\dimexpr\paperwidth-20cm\relax,-0.5cm){\parbox[t]{19cm}{\copyright 2023 IEEE. Personal use of this material is permitted. Permission from IEEE must be obtained for all other uses, in any current or future media, including reprinting/republishing this material for advertising or promotional purposes, creating new collective works, for resale or redistribution to servers or lists, or reuse of any copyrighted component of this work in other works. DOI: \url{https://doi.org/10.1109/ICASSP48485.2024.10446445}}}%

}}

\def\Fig#1{{ Fig.\ \ref{#1}}}
\def\Figu#1{{ Figure \ref{#1}}}

\def\Tabl#1{{ Table \ref{#1}}}

\title{Enhanced Color Palette Modeling for Lossless Screen Content Compression}
\name{Hannah Och$^{\star}$ \qquad  Shabhrish Reddy Uddehal$^{\dagger \star}$ \qquad Tilo Strutz$^{\dagger}$ \qquad Andr\'{e} Kaup$^{\star}$\thanks{
		This work has been funded by the Deutsche Forschungsgemeinschaft (DFG, German Research Foundation) - project ID 438221930.
}}
\address{$^{\star}$Multimedia Communications and Signal Processing\\Friedrich-Alexander Universität Erlangen-Nürnberg (FAU), Erlangen, Germany \\
	$^{\dagger}$Department of Electrical Engineering and Computer Science\\
		Coburg University of Applied Sciences and Arts, Coburg, Germany}
\begin{document}
\ninept
\maketitle
\begin{abstract}
Soft context formation is a lossless image coding method for screen content. It encodes images pixel by pixel via arithmetic coding by collecting statistics for probability distribution estimation. Its main pipeline includes three stages, namely a context model based stage, a color palette stage and a residual coding stage. Each subsequent stage is only employed if the previous stage can not be applied since necessary statistics, e.g. colors or contexts, have not been learned yet. We propose the following enhancements: First, information from previous stages is used to remove redundant color palette entries and prediction errors in subsequent stages. Additionally, implicitly known stage decision signals are no longer explicitly transmitted. These enhancements lead to an average bit rate decrease of $1.07\%$ on the evaluated data. Compared to VVC and HEVC, the proposed method needs roughly 0.44 and 0.17 bits per pixel less on average for 24-bit screen content images, respectively.
\end{abstract}
\section{Introduction}
Screen content images (SCIs) refer to images that can typically be seen on computer desktops, smartphones or similar devices. In contrast to camera-captured content, they often contain large uniform areas, a limited number of unique colors, and repeating image blocks such as repeated letters on a text document. Such characteristics can be exploited for image compression, which is a very relevant topic considering the increased usage of applications such as e-learning, video conferencing and remote desktop sharing. Current state-of-the-art image and video compression standards address this issue by dedicated screen content coding tools. The High Efficiency Video Coding (HEVC) standard \cite{Sul12} as well as the Versatile Video Coding (VVC) standard \cite{Bro21} both incorporate, for example, the coding tools intra block copy (IBC), palette mode, and adaptive color transform with slight differences in implementation \cite{Xu16,VVCSCC}. VVC contains additional tools, such as transform skip residual coding \cite{VVCTSRC} and block-based differential pulse-code modulation \cite{VVCBDPCM}, which help in further improving coding efficiency for screen content.

Next to these block-based codecs, methods based on ideal entropy coding and probability distribution modeling have been shown to work very well for lossless screen content compression. The free lossless image format (FLIF) \cite{Sne16} is designed for a wide range of image types. Pixel colors are predicted from direct neighbors and the error is encoded using entropy coding. Vanilc, proposed in \cite{WeiAm16}, can handle lossless coding of volumetric medical data as well as non-natural images and incorporates histogram sparsification specifically for those images not generated by image sensors. JPEG-LS \cite{JPEG_LS} is another well-known algorithm based on pixel by pixel prediction in conjunction with Rice codes. A recent state-of-the-art lossless coder for SCIs uses the Soft Context Formation (SCF) \cite{Str19}. During encoding, the statistics of the image are learned and used for probability estimation and entropy coding of the pixel values. This approach works very well for SCIs with a limited number of colors and repeated image areas. It outperforms HEVC 
in the lossless case for images with less than 90000 colors in most cases. Additional modifications in \cite{Och21} and \cite{Udd23} have further improved the coding efficiency, especially for images with more colors. 

The core of the SCF coder is the soft context formation, which exploits patterns of neighboring pixels as context for estimation of color probabilities. Given the probability distribution, the color of the pixel is encoded using arithmetic coding. This is referred to as 'Stage 1'. However, since it may happen that the statistics for a given context or the current color are not learned yet, other methods have to be used in these cases. Stage 2 uses a global color palette for probability distribution estimation and for colors not yet part of the color palette (i.e., not seen before) Stage 3 must encode the pixel via prediction and residual coding. 
Stage 1 is by far the most efficient one in terms of bit rate, Stage 3 is the least efficient one.
In this paper, we aim to increase the coding efficiency of the SCF coder by exploiting information that can be derived from previous stages. This either improves the probability modeling or simply reduces the number of syntax elements to be transmitted like stage decision flags.

In the following, we first describe the SCF coder in more detail. We give special focus to the components important in the following explanations. Afterwards, we explain the proposed enhancements for redundancy reduction. Finally, we evaluate the effect of the changes on multiple SCI data sets and analyze the gains in detail.

\section{Review of the Soft Context Formation Coder}
\label{SCF}
	\begin{figure}
		\hfil\includegraphics[scale=0.19]{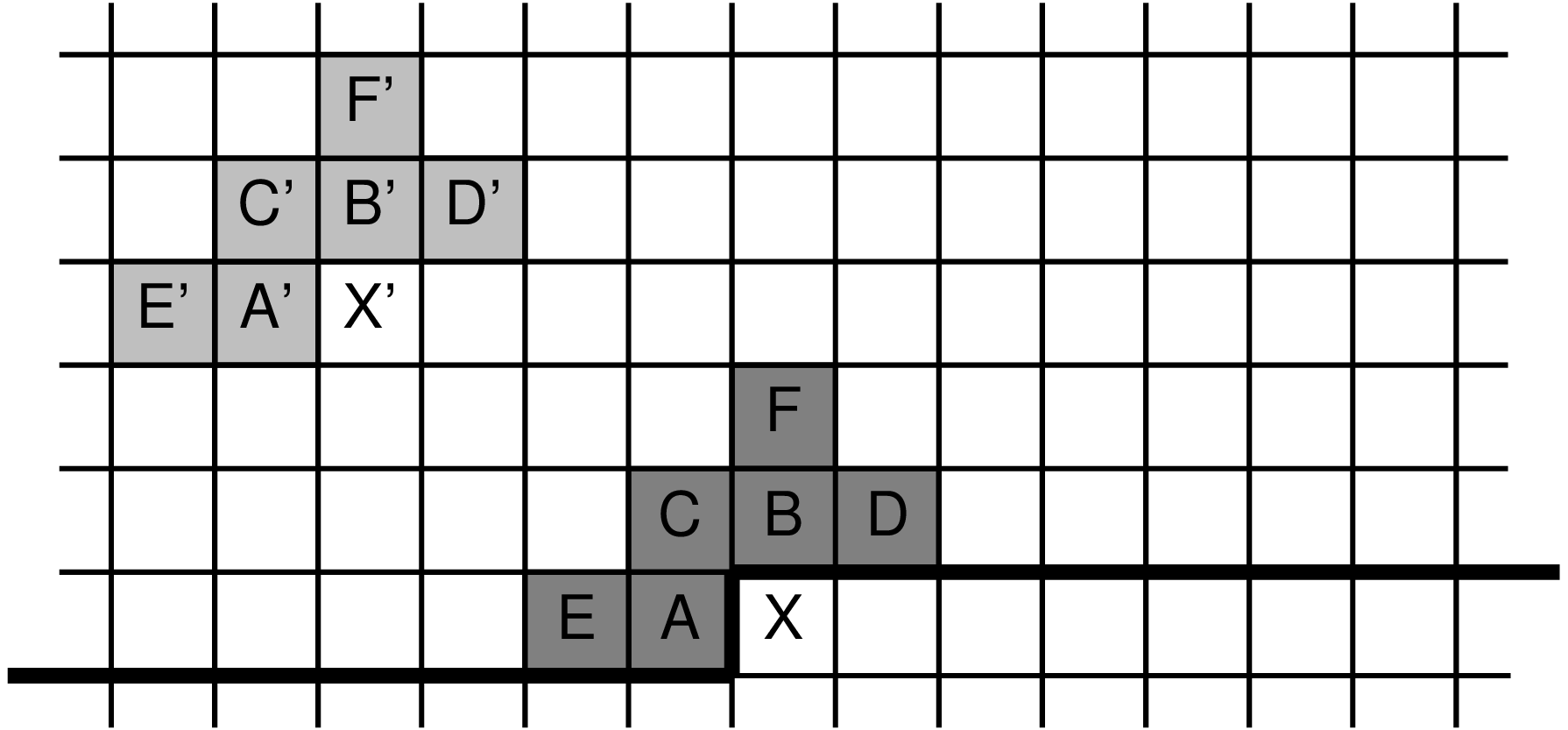}
			\vskip -0.6em
		\caption{\label{fig_template_matching}Context pattern: If the values in the template (A, B, \dots , F) are similar to (A', B', \dots, F'), then the current value at position X is likely to be similar to the value at X'.}
			\vskip -0.7em
	\end{figure}
The SCF coder processes an image pixel by pixel. First, based on six causal neighboring pixels as visualized in \Fig{fig_template_matching}, a probability distribution of possible associated colors is estimated. To this end, during the coding process for each distinct pattern a distribution of all associated colors $x$ is updated. It should be noted that $x$ indicates an RGB color with all its channels. The color distributions of similar patterns to the current one are then merged to create the distribution of possible colors for the current pixel. For more information on the similarity criterion, please refer to \cite{Str16b}. In the following, we will call this process Stage 1 and the distribution of colors based on pattern matching `Stage 1 distribution'. This distribution contains counters for colors $\{x_i|i \in \mathcal{S}\}$, where the set $\mathcal{S}$ contains the indices of all colors that are part of the current Stage 1 distribution. If the current color $x$ is part of this distribution, it can be directly encoded using arithmetic coding. If this is not the case, meaning no similar patterns were found or the color is not yet part of the associated distribution, an escape signal is sent and Stage 2 is initiated. 

In Stage 2, the color can be encoded based on a color palette that contains counters for all previously seen colors. Since it is possible that a completely new color is occurring, first a flag signals whether the color is already part of the color palette and Stage 2 is applicable. If yes, the distribution of the colors $\{x_i|i \in \mathcal{P}\}$ will be used for arithmetic encoding of the entire RGB color, where the set $\mathcal{P}$ contains all colors in the current color palette. Since images with many colors will have a large color palette and thus lead to broad probability distributions, some enhancements to this stage have been integrated in \cite{Str19}. Based on prediction errors in the neighborhood a radius is calculated. Then, the color palette is split in two sub-palettes: One sub-palette contains all colors $\{x_i|i \in \mathcal{P}_\mathrm{r} \}$ that are inside the radius of the predicted value $\hat{x}$. The other sub-palette contains the rest of the known colors $\{x_i|i \in \mathcal{P}\setminus \mathcal{P}_\mathrm{r} \} $. A flag indicates to which sub-palette the current color belongs. Afterwards, the counts of the colors in the corresponding sub-palette are used to estimate the distribution and to encode the pixel arithmetically. 

\begin{figure}
\centering
\hfil\input{Images/diagram_neu_110621.tex}
	\vskip -0.6em
\caption{\label{stages}Block diagram of SCF method for one pixel: If the color $x$ has already happened in conjunction with a similar pattern, it is coded in Stage 1. Otherwise, if the color already appeared in the image, it is coded via palette-based coding. If neither case has occurred, it is encoded using residual coding. Finally, the histograms and the pattern list are updated.}
	\vskip -0.6em
\end{figure}
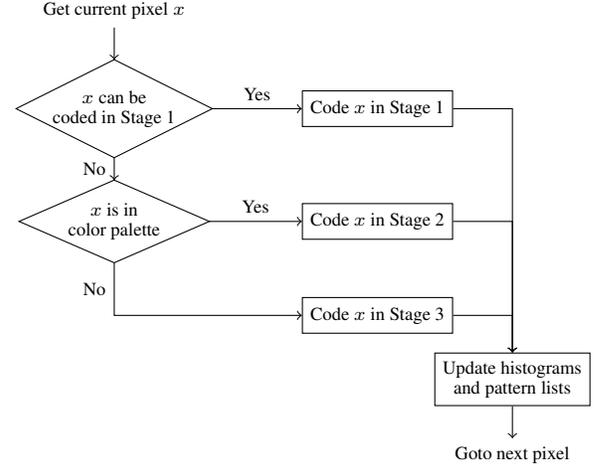

Finally, Stage 3 is used for colors appearing for the first time. The current pixel is predicted using an enhanced median adaptive predictor \cite{Bed04,Str16b}. The prediction error is encoded based on a histogram of prediction errors for each color channel consecutively. More details on Stage 3 can be found in \cite{Och21}. An overview of the SCF processing with all its stages is visualized in \Fig{stages}.

\section{Improvements to the Soft Context Formation Coder}
In the following, we will describe modifications to the SCF coder in Stage 2 and 3, which remove redundancy by utilizing already known information from Stage 1, Stage 2 or the SCF header.

\subsection{Enhanced Color Palette Modeling}
\label{palettemodeling}
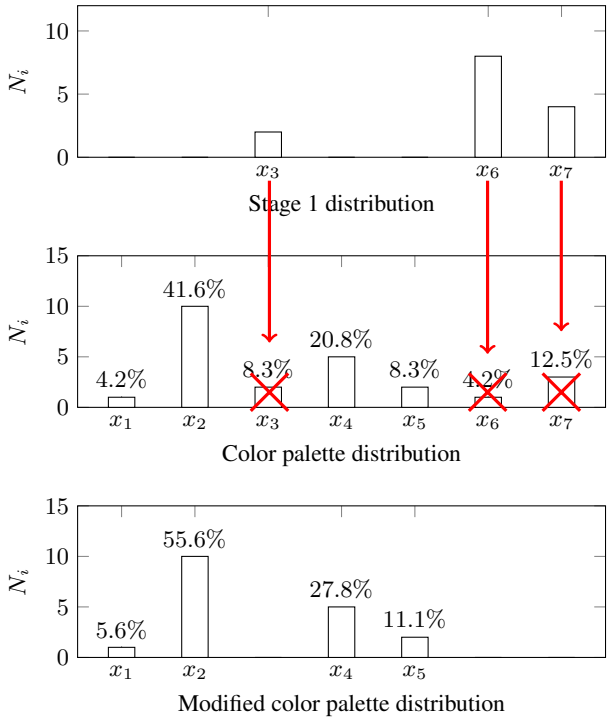
\begin{figure}
	\centering´
	\hfil\input{Visualization_ICASSP24/Visualization_s1.tex} 
	\vskip -0.6em
	\hfil\input{Visualization_ICASSP24/Visualization_palette.tex}
	\vskip -0.6em
	\hfil\input{Visualization_ICASSP24/Visualization_Modpalette.tex}
	\input{Visualization_ICASSP24/overlay.tex}
	\vskip -0.6em
	\caption{\label{distributions}Example of color palette reduction with rounded probability estimates for Stage 2 coding.}
\vskip -0.7em
\end{figure}

The first modifications use the distributions from previous stages by removing values, i.e. color triples, from the current stage distribution, if they have already been ruled out by the previous stage. Let us first look at Stage 2: If Stage 2 is needed, necessarily Stage 1 was not applicable for encoding the color of the current pixel. This means that none of the colors included in the Stage 1 distribution $\{x_i|i \in \mathcal{S}\}$ can be the color $x$ of the current pixel. However, the colors in the Stage 1 distribution are a subset of the color palette distribution, meaning $\mathcal{S} \subseteq \mathcal{P}$. Thus, the distribution used for encoding a color in Stage 2 may contain colors with a non-zero probability that we already know to be impossible and can therefore be removed from the distribution. The proposed process implemented in the SCF coder is visualized exemplarily in \Figu{distributions}. The colors $x_3$, $x_6$ and $x_7$, which are part of the Stage 1 distribution, can be excluded from the color palette histogram used in Stage 2, so that a modified color palette with fewer entries is achieved.

To show the result of this modification mathematically, we leave out the color palette splitting in the following equations for easier description, however, the math is the same. Let $N_{i}$ be the count of the $i$-th color in the color palette and $N_\mathrm{p} = \sum_{i \in \mathcal{P}}N_{i}$ the summed counts of all colors in the color palette. Then, the probability of the $i$-th color $x_i$ in the color palette is given as 
\begin{equation}
	p(x_{i}) = \dfrac{N_i}{N_\mathrm{p}} \,\,\,\,\,\,\forall \, i \in \mathcal{P}.
\end{equation}
The colors $\{x_i|i \in \mathcal{S} \}$ from the Stage 1 distribution cannot be the current color $x$. As such, they can be removed from the palette distribution, so that the total sum of counts in the modified palette histogram is given as  
\begin{equation}
	N_\mathrm{p}^* = \sum_{i \in \mathcal{P} \setminus \mathcal{S}}N_{i} \leq N_\mathrm{p}.
\end{equation}
Consequently, for all probabilities of the possible colors $\{x_i|i \in \mathcal{P} \setminus \mathcal{S} \}$ it holds that
\begin{equation}
	p(x_{i})^* = \dfrac{N_i}{N_\mathrm{p}^*} \geq \dfrac{N_i}{N_\mathrm{p}} = p(x_{i} ) \,\,\,\,\,\,\forall\, i \in \mathcal{P} \setminus \mathcal{S}.
\end{equation}
The bit cost for arithmetic coding of a value based on a probability distribution is given by its self-information. Since the logarithm is a monotonously increasing function and has its only zero-crossing at 1, we get a smaller or equal bit cost for the new probability according to
\begin{equation}
	I(x_{i})^*=-\log_2{p(x_{i})^*} \leq I(x_{i} )=-\log_2{p(x_{i})} \,\,\forall\, i \in \mathcal{P} \setminus \mathcal{S}.
\end{equation}

Thus, the bits needed to encode a color based on the color palette will be  less or equal after first removing all Stage 1 distribution colors from the color palette. 

Similarly, if a color has to be encoded in Stage 3, the colors that are part of the color palette cannot be the color of the current pixel. We can utilize this information to remove redundancy from Stage 3 using the same principle as for Stage 2. However, two points have to be considered.

First, each channel value is coded consecutively in Stage 3. This means that, for the coding of the first two channels (in the case of an RGB image), we cannot exclude all known values from the distribution. As an example, let the color $x = (0,1,2)^\top$ be already part of the color palette and thus not a viable color for a Stage 3 pixel. When we process the values of the first two channels, $x = (0,1,x_k)^\top$ is still a possible triplet, if, e.g., the current pixel would have the color $(0,1,3)^\top$. Thus, values can be excluded from the distributions only once all channels except the last have been  coded. 

Second, in Stage 3, prediction errors are encoded instead of RGB colors. As such, all colors included in the color palette that we want to exclude from the Stage 3 distribution have to be transformed into their corresponding prediction error $e_{i,k}$ via 
\begin{equation}
	e_{i,k} = \hat{x}_k - x_{i,k},
\end{equation}
where $\hat{x}_k$ is the predicted value of the final channel $k$ of the current pixel and $x_{i,k}$ the final channel of the $i$-th color that we want to remove from the Stage 3 distribution. 

Taking these two points into account, we can apply the same process as before, and remove the prediction errors of the final channel of all colors $\{x_i|i \in \mathcal{P}\}$ from the  prediction error histograms in Stage 3. As shown above, this will directly lead to less or equal number of bits needed.

\subsection{Efficient Stage Decision Signaling}
Finally, two additional modifications lead to fewer transmissions of coding decisions in Stage 2. One flag is needed to indicate whether a color is part of the color palette and thus will be encoded in Stage 2. Currently, this flag is transmitted for all pixels that are not directly encoded in Stage 1. However, this is not necessary in certain conditions. The header of the SCF contains the total number of unique colors in the image. Thus, without increase in computational complexity and without additional information, we can adapt the SCF coder as follows: During coding, the number of colors in the color palette is counted. As soon as all unique colors in an image have been found, i.e. the total number of colors transmitted in the header is equal to the number of colors in the color palette, no further new colors can appear. This means that all future pixels can be coded at the latest in Stage 2 and Stage 3 is no longer needed. Thus, this flag has to be true for all future pixels and no further bits have to be allocated in the following image for this decision. 

Additionally, as mentioned in Section \ref{SCF}, the color palette in Stage 2 is split in two sub-palettes. A flag indicates which sub-palette is used. However, this flag does not always have to be transmitted explicitly. As explained in Section \ref{SCF} the sub-palette partition is based on the predicted value $\hat{x}$ of the current pixel and a radius, which are both independent of the content of the color palette. As such, it is possible that all available colors are part of either $\mathcal{P}_\mathrm{r}$ or $\mathcal{P} \setminus \mathcal{P}_\mathrm{r}$. If that is the case, this flag can safely be skipped since both encoder and decoder know which sub-palette has to be the correct one without explicit signaling.

\section{Evaluation}
To evaluate the effect of the proposed enhancements, we use multiple SCI data sets with varied content. The data set HEVC-CTC contains 14 single frames taken from HEVC CTC test sequences as listed in  in \cite{Str19} for \textit{Test set 2}. The sequences are from the common test conditions for screen content coding \cite{HEVC_SCC_CTC} and screen content range extension experiments \cite{HEVC_RCE3}. Additionally, we use the well-known SIQAD data set \cite{Hua15} with 20 images and the SCID data set \cite{Zha17, SCID} with 50 images. Data sets SC-Text and SC-Mixed contain screen shots of webpages with a resolution of $1360\times 768$ from the `text' and `mixed' stimuli of \cite{She14}, respectively. All data sets contain RGB color images with an uncompressed bit rate of 24 bits per pixel.  In total, we evaluate our proposed modifications on 173 diverse SCIs.
Since the method produces bit-exact reconstructions of images, there is no need to investigate image quality, and we will focus on bit rate in the following. 																																												 
\begin{table*}[t]
	\centering
	\caption{\label{comparison}Compression performances of the proposed SCF method. The table lists the bit rate in bit per pixel averaged over each RGB color data set as well as relative bit rates in comparison to the reference method \cite{Och21}.}
	\vskip -0.6em
	\hfil
	{
		\input{Images/compression_performance.tex}

	}
	\vskip -0.6em
\end{table*}

\Tabl{comparison} shows the bit rates of the images averaged for each data set separately as well as all data sets combined after compression using the SCF coder with and without the proposed modifications. As the base SCF, we use the version from \cite{Och21}, which is the original SCF \cite{Str16b} with various enhancements as explained in \cite{Str19} as well as improved exception handling \cite{Str20} and residual coding \cite{Och21}. To break down the savings to the different modifications, each adaptation is successively added, i.e. SCF Base corresponds to \cite{Och21}, Base~+~P applies the reduced color palette in Stage 2 and Base~+~PR additionally the improvements to the residual coding stage. Finally, Base + PRF contains all proposed modifications including the removal of redundant flag signaling. To compare against other state-of-the-art methods, HEVC, VVC,
and FLIF results are added as well. For HEVC, we use the reference codec HM 16.21 + SCM 8.8 with all-intra configuration according to the common test conditions (CTC) for screen content given in \cite{HEVC_SCC_CTC}.
For VVC, we compress the images using the reference codec VTM 17.2 with all-intra and the configurations for RGB 4:4:4 format and non-camera captured content according to \cite{Chao2020}. 
This means, that both HM and VTM use all available screen content specific coding tools, such as IBC or palette mode, in our evaluation when appropriate. HM and VTM are applied in the lossless mode on the original data in RGB 4:4:4. 

It is immediately noticeable that VTM has worse results than HM. Though this is unexpected, this result corresponds to \cite{T0111}, where the lossless coding mode of the VTM is investigated for YCgCo-R and RGB. Amongst other findings, it is shown that VTM 10.0 has a worse compression efficiency than the HM with its screen content coding extensions in lossless mode for RGB 4:4:4 images. 

As we can see, the base SCF is already better than HEVC, VVC, and FLIF for all data sets. FLIF needs roughly $14\%$ more bit rate on average, VVC $18\%$ and HEVC roughly $6\%$. We reduce the needed bit rate by additional $0.49\%$ compared to SCF Base using the color palette modification described in Section \ref{palettemodeling}. The highest gains are achieved for test set HEVC-CTC with $0.56\%$ bit rate savings and the lowest for SIQAD with $0.42\%$. The modification of the residual coding stage achieves further $0.55\%$ savings on average. Here, the gains are a bit more varied between $0.78\%$ for SIQAD and $0.17\%$ for HEVC-CTC. Finally, the removal of the redundant flags leads to  $0.03\%$ gain. In total, $1.07\%$ bit rate is saved on average. With all modifications, the bit rate savings achieved range from $0.28\%$ to $2.83\%$ for single images in the given data set. It is important to note that the proposed modifications cannot lead to worse coding efficiency in terms of bit rate for any image, since we only remove redundancies, i.e., all images can be compressed better than before.

\begin{table}[t]
		\centering
	\caption{\label{timeComparison}Accumulated encoding and decoding time over all test images in seconds.}
	\vskip -0.6em
	\hfil
	{\setlength{\tabcolsep}{2pt}
		\input{Images/time_performance.tex}

	}
	\vskip -0.7em
\end{table}
Finally, the effects of the modifications on processing time are considered. It should be noted that the SCF coder is not optimized for complexity or speed. For evaluation of the coding time, we run all codecs on the same device (Intel Core i7-10510U CPU processor) four times over all 173 images and average the accumulated encoding and decoding time. \Tabl{timeComparison} shows the results. It is noticeable, that there is a slight increase in encoding time when using the proposed modifications as expected. However, since all necessary information for the modifications applicable in Stage 2, namely color palette reduction and flag removal, can be collected in one search through the color palette, the corresponding increase in complexity is low. The main contributors to the increase in coding time are the additional checks for removable symbols in the residual coding stage. The increase in complexity could be reduced with further optimization, such as a more sophisticated palette search in comparison to brute force. The decoding time is even slightly lower than before. This is purely due to a change in implementation where the decoder can save itself a search through the color palette, as all information required for decoding in Stage 2 is already gathered in the one-time palette search described above, which is implemented as two palette searches in \cite{Och21}. In terms of complexity, the SCF needs roughly $60\%$ of the encoding time of VTM. However, it is not yet competitive with the HM on average. Since the SCF is a  symmetric coder, this is especially true for decoding. However, the coding time for SCF is largely dependent on the number of colors per image, i.e. the more colors an image contains, the larger the pattern lists and color palette, and the higher the coding time. For images with very few colors, the encoding time needed by SCF and HM are in the same order of magnitude, such as `SCI05' from SCID with 6687 unique colors, which is encoded by SCF in 23 seconds and by HM in 18 seconds.

\section{Summary and Outlook}
The SCF Coder is a lossless coder for SCIs. Due to the hierarchical structure of the SCF coder and the separate handling of the different stages, there are some redundancies in the coding process of the previous versions that we can exploit. We propose modifications to the codec that remove these redundancies without any need for additional side information. In Stage 2 and Stage 3, color histograms from previous stages are utilized to improve the probability distributions used for encoding of the current pixel color. Additionally, redundant Stage 2 decision flags are no longer explicitly transmitted but deduced from available information instead. On the evaluated data sets, we gain $1.07\%$ rate savings an average, most of which stem from the removal of redundant symbols from the probability models when encoding color triplets in Stage 2 or the residual in Stage 3. When compared to VVC and HEVC, the proposed method saves 0.44 and 0.17 bits per pixel on average, respectively.

\balance
\bibliographystyle{IEEEbib}
\bibliography{literature}

\end{document}

%% file: Images/diagram_neu_110621.tex
\usetikzlibrary{shapes.geometric}
\usetikzlibrary{positioning}

\tikzstyle{block} = [draw, fill=white, rectangle, minimum height=1.5em, minimum width=5em, align=center]
\tikzstyle{decision} = [draw, fill=white,diamond, aspect=2,minimum height=2em, minimum width=8em,align=center, inner sep=0pt, outer sep=0pt]
\tikzstyle{input} = [coordinate]
\tikzstyle{output} = [coordinate]
\tikzstyle{pinstyle} = [pin edge={to-,thin,black}]
\tikzstyle{pinstyle2} = [pin edge={-to,thin,black}]

\tikzset{font=\scriptsize}
\begin{tikzpicture}
	\noindent
    \node [decision, pin={[pinstyle]above:Get current pixel $x$}, node distance=1.5cm] (stage1decision){$x$ can be\\ coded in Stage 1};
		\node [decision,below of=stage1decision, node distance=1.5cm] (stage2decision){$x$ is in\\ color palette};
		\node [block, right of=stage1decision, node distance=3.5cm] (stage1){Code $x$ in Stage 1};
		\node [block, right of=stage2decision, node distance=3.5cm] (stage2){Code $x$ in Stage 2};
		\node [block, below of=stage2, node distance=1.25cm] (stage3){Code $x$ in Stage 3};
		\node [block, below right=0.25cm and -0.25cm of stage3,pin={[pinstyle2]below:Goto next pixel}] (update){Update histograms\\and pattern lists};
		
		\draw  [->] (stage1decision) -- node [name=Yes1, midway, above] {Yes} (stage1);
		\draw  [->] (stage2decision) -- node [name=Yes2, midway, above] {Yes} (stage2);
		\draw  [->] (stage1decision) -- node [name=No1, midway, left] {No} (stage2decision);
		\draw [->] (stage2decision) |- node [name=No2, near start, left] {No} (stage3);
		\draw [->] (stage1) -| (update);
		\draw [->] (stage2) -| (update);
		\draw [->] (stage3) -| (update);
\end{tikzpicture}

%% file: Visualization_ICASSP24/Visualization_s1.tex

\def\mywidth{\columnwidth}
\begin{tikzpicture}[remember picture]
	\begin{axis}[width=\mywidth,height=\mywidth/2.39,
		symbolic x coords={$x_1$,$x_2$,$x_3$,$x_4$,$x_5$,$x_6$,$x_7$},
		ylabel = {$N_i$},
		xlabel = {Stage 1 distribution},
		ymin=0,
		ymax=12,
		xtick={$x_3$,$x_6$,$x_7$}]
		\addplot[ybar,fill=white] coordinates {
			($x_1$,0)
			($x_2$,0)
			($x_3$,2)
			($x_4$,0)
			($x_5$,0)
			($x_6$,8)
			($x_7$,4)
		};
	\end{axis} 

\node (x3_s1) at (2.55,-0.2) {};
\node (x7_s1) at (6.43,-0.2) {};
\node (x6_s1) at (5.45,-0.2) {};
\end{tikzpicture}
	

%% file: Visualization_ICASSP24/Visualization_palette.tex

\def\mywidth{\columnwidth}
\newcommand{\Cross}{$\mathbin{\tikz [x=3.5ex,y=3.5ex,line width=.3ex, red] \draw (0,0) -- (1,1) (0,1) -- (1,0);}$}%
\begin{tikzpicture}[remember picture]
	\begin{axis}[width=\mywidth,height=\mywidth/2.39,
		symbolic x coords={$x_1$,$x_2$,$x_3$,$x_4$,$x_5$,$x_6$,$x_7$},
		ylabel = {$N_i$},
		xlabel = {Color palette distribution},
		ymin=0,
		ymax=15,
				 nodes near coords,
		xtick=data,
				point meta=explicit symbolic]
		\addplot[ybar,fill=white] coordinates {
			($x_1$,1)[$4.2\%$]
			($x_2$,10)[$41.6\%$]
			($x_3$,2)[$8.3\%$]
			($x_4$,5)[$20.8\%$]
			($x_5$,2)[$8.3\%$]
			($x_6$,1)[$4.2\%$]
			($x_7$,3)[$12.5\%$]
		};
	\end{axis} 

\node (x3_p1) at (2.55,0.75) {};
\node (x7_p1) at (6.43,0.9) {};
\node (x6_p1) at (5.45,0.6) {};

\node (x3_p11) at (2.55,0.2) {\Cross};
\node (x7_p11) at (6.43,0.2) {\Cross};
\node (x6_p11) at (5.45,0.2) {\Cross};
\end{tikzpicture}
	

%% file: Visualization_ICASSP24/Visualization_Modpalette.tex

\def\mywidth{\columnwidth}
\begin{tikzpicture}[remember picture]
	\begin{axis}[width=\mywidth,height=\mywidth/2.39,
		symbolic x coords={$x_1$,$x_2$,$x_3$,$x_4$,$x_5$,$x_6$,$x_7$},
		ylabel = {$N_i$},
		xlabel = {Modified color palette distribution},
		ymin=0,
		ymax=15,
		 nodes near coords,
		xtick={$x_1$,$x_2$,$x_4$,$x_5$},
		point meta=explicit symbolic]
		\addplot[ybar,fill=white] coordinates {
			($x_1$,1)[$5.6\%$]
			($x_2$,10)[$55.6\%$]
			($x_3$,0)
			($x_4$,5)[$27.8\%$]
			($x_5$,2)[$11.1\%$]
			($x_6$,0)
			($x_7$,0)
		};

	\end{axis} 

\node (x3_p2) at (2.675,0.2) {};
\node (x7_p2) at (6.775,0.2) {};
\node (x6_p2) at (5.775,0.2) {};


\end{tikzpicture}
	

%% file: Visualization_ICASSP24/overlay.tex

\def\mywidth{\columnwidth}

\begin{tikzpicture}[remember picture,overlay]
	\path[->,red,very thick] (x3_s1) edge (x3_p1);
	\path[->,red,very thick] (x6_s1) edge (x6_p1);
	\path[->,red,very thick] (x7_s1) edge (x7_p1);
\end{tikzpicture}
	

%% file: Images/compression_performance.tex
\begin{tabular}{|l|c||c|c|c|c|c|c|c|}
	\hline
	& Num. of	& 	 				& 									& HM-16.21	       	& \multicolumn{4}{c|}{SCF}                                                                                                                              \\
	& images  	&	FLIF			&	VTM 17.2					& SCM-8.8     	   	& \multicolumn{1}{c|}{Base \cite{Och21}} 	& \multicolumn{1}{c|}{Base + P}    		& \multicolumn{1}{c|}{Base + PR}  			& Base + PRF            \\ \hline \hline
	HEVC-CTC \cite{HEVC_SCC_CTC,HEVC_RCE3}& 14  & 2.019			& 2.549			& 2.129          	& \multicolumn{1}{c|}{1.944}           	& \multicolumn{1}{c|}{1.933}         	& \multicolumn{1}{c|}{1.930}       		& 1.928         \\
	\textit{Percentage} 		&     		& \textit{103.87\%}	& \textit{131.17\%}	& \textit{109.55\%} 	& \multicolumn{1}{c|}{\textit{100.00\%}} 	& \multicolumn{1}{c|}{\textit{99.44\%}} & \multicolumn{1}{c|}{\textit{99.27\%}} 	& \textit{99.21\%} \\ \hline
	SIQAD \cite{Hua15}      & 20 	 		& 4.104			& 4.423			& 3.988          	& \multicolumn{1}{c|}{3.750}           	& \multicolumn{1}{c|}{3.734}         	& \multicolumn{1}{c|}{3.705}       		& 3.704         \\
	\textit{Percentage} 		&     		& \textit{109.45\%}	& \textit{117.97\%}	& \textit{106.34\%} 	& \multicolumn{1}{c|}{\textit{100.00\%}} 	& \multicolumn{1}{c|}{\textit{99.58\%}} & \multicolumn{1}{c|}{\textit{98.80\%}} 	& \textit{98.79\%} \\ \hline
	SCID \cite{Zha17}        & 40  		& 3.438			& 3.7345	& 3.373       	& \multicolumn{1}{c|}{3.202}           	& \multicolumn{1}{c|}{3.185}        	& \multicolumn{1}{c|}{3.162}      			& 3.161        \\
	\textit{Percentage} 		&     		& \textit{107.36\%}	& \textit{116.63\%}	& \textit{105.33\%} 	& \multicolumn{1}{c|}{\textit{100.00\%}} 	& \multicolumn{1}{c|}{\textit{99.46\%}} & \multicolumn{1}{c|}{\textit{98.74\%}} 	& \textit{98.73\%} \\ \hline
	SC-Text	\cite{She14}		& 50			&	1.859			& 1.732			& 1.535 			& 1.428 									& 1.421								& 1.416									& 1.415 \\
	\textit{Percentage} 		&     		& \textit{130.16\%}	& \textit{121.28\%}	& \textit{107.47\%} 	& \multicolumn{1}{c|}{\textit{100.00\%}} 	& \multicolumn{1}{c|}{\textit{99.52\%}} & \multicolumn{1}{c|}{\textit{99.15\%}} 	& \textit{99.11\%} \\ \hline
	SC-Mixed \cite{She14}		& 49			& 2.199			& 2.152			& 1.975 			& 1.869 									& 1.861 								& 1.853									& 1.853 \\
	\textit{Percentage} 		&     		& \textit{117.68\%}	& \textit{115.12\%}	& \textit{105.65\%} 	& \multicolumn{1}{c|}{\textit{100.00\%}} 	& \multicolumn{1}{c|}{\textit{99.56\%}} & \multicolumn{1}{c|}{\textit{99.14\%}} 	& \textit{99.11\%} \\ \hline \hline
	Total               		& 173 		& 2.593		& 2.691			& 2.416         	& \multicolumn{1}{c|}{2.273}           	& \multicolumn{1}{c|}{2.262}        	& \multicolumn{1}{c|}{2.250}       		&  2.249				\\
	\textit{Percentage} 		&	   			& \textit{114.06\%}	& \textit{118.38\%}	& \textit{106.28\%} 	& \multicolumn{1}{c|}{\textit{100.00\%}} 	& \multicolumn{1}{c|}{\textit{99.51\%}} & \multicolumn{1}{c|}{\textit{98.96\%}} 	& \textit{98.93\%} \\ \hline
\end{tabular}

%% file: Images/time_performance.tex
\begin{tabular}{|c||c|c|c|c|}
\hline
					&	& HM-16.21				& \multicolumn{2}{c|}{SCF}          \\
                  &  VTM 17.2 & SCM-8.8 & Base \cite{Och21}	& Base + PRF            \\ \hline \hline
Encoding time [s]	& 55316 & 4510	& 31409    			  			& 32990   \\
Decoding time [s]	& 45 & 32 &    29184		& 27411  \\ \hline
\end{tabular}